\documentclass[3p]{elsarticle}

\usepackage{amssymb}

\usepackage{graphicx}
\usepackage{balance}
\usepackage{color}
\usepackage{xcolor}
\usepackage{multirow}
\usepackage{caption}
\usepackage{enumitem}

\usepackage{subcaption,epsfig,amsfonts,amsmath,cases,amssymb,graphics, latexsym, times, algorithmic, algorithm, bbm, color, soul,wrapfig,comment}

\journal{Smart Health}

\begin{document}

\begin{frontmatter}



\title{mTOCS: Mobile Teleophthalmology in Community Settings to improve Eye-health in Diabetic Population}


\author[mucs]{Jannat Tumpa}
\ead{jannat.tumpa@marquette.edu}

\author[mucs]{Riddhiman Adib}
\ead{riddhiman.adib@marquette.edu}

\author[mucs]{Dipranjan Das}
\ead{dipranjan.das@marquette.edu}

\author[mcw]{Nathalie Abenoza}
\ead{nabenoza@mcw.edu}

\author[mcw]{Andrew Zolot}
\ead{azolot@mcw.edu}

\author[mcw]{Velinka Medic}
\ead{vmedic@mcw.edu}

\author[mcw]{Judy Kim}
\ead{jekim@mcw.edu}

\author[ucc]{Al Castro}
\ead{acastro@unitedcc.org}

\author[ucc]{Mirtha Sosa Pacheco}
\ead{msosa@unitedcc.org}


\author[mhd]{Jay Romant}
\ead{jaroman@milwaukee.gov}

\author[mucs]{Sheikh Iqbal Ahamed}
\ead{sheikh.ahamed@marquette.edu}

\address[mcw]{Medical College of Wisconsin, Milwaukee, Wisconsin, United States}

\address[ucc]{United Community Center, Milwaukee, Wisconsin, United States}

\address[mhd]{City of Milwaukee Health Department, Milwaukee, Wisconsin, United States}

\address[mucs]{Department of Computer Science, Marquette University, Wisconsin, United States}


\begin{abstract}
Diabetic eye diseases, particularly Diabetic Retinopathy,is the leading cause of vision loss worldwide and can be prevented by early diagnosis through annual eye-screenings. However, cost, healthcare disparities, cultural limitations, etc. are the main barriers against regular screening. Eye-screenings conducted in community events with native-speaking staffs can facilitate regular check-up and development of awareness among underprivileged communities compared to traditional clinical settings. However, there are not sufficient technology support for carrying out the screenings in community settings with collaboration from community partners using native languages. In this paper, we have proposed and discussed the development of our software framework, "Mobile Teleophthalomogy in Community Settings (mTOCS)", that connects the community partners with eye-specialists and the Health Department staffs of respective cities to expedite this screening process. Moreover, we have presented the analysis from our study on the acceptance of community-based screening methods among the community participants as well as on the effectiveness of mTOCS among the community partners. The results have evinced that mTOCS has been capable of providing an improved rate of eye-screenings and better health outcomes. 

\end{abstract}


\end{frontmatter}

\section{Introduction}

Diabetes poses a major threat to human health, and according to the World Health Organization, it is one of the leading causes of death worldwide \cite{diabetes_who}. About 422 million people around the globe, particularly in low and middle-income countries, currently have diabetes, and this number is increasing rapidly \cite{cho2018idf}. In the context of the U.S., the number of Americans with diabetes is projected to be doubled by 2050 \cite{cdc2010}. People suffering from diabetes for a prolonged period are at higher risk of developing retinal complications that can eventually cause vision loss \cite{yau2012global}. Diabetic Retinopathy (DR) is one such retinal complication which is prevalent in 90\% people who have Type 1 diabetes for a more extended period. However, it can be prevented to a great extent through early-diagnosis and intervention \cite{chew2014effects}.

Few racially and ethnically minority communities are disproportionately affected by DR \cite{barsegian2017diabetic}. Centers for Disease Control and Prevention (CDC) has published several statistical analysis, stating that, Latinos and African Americans are at particularly high risk for Diabetic Retinopathy, and they also have the lowest record of having recommended annual eye exams \cite{cdcvision} due to different health disparities prevalent in U.S. \cite{zambelli2012disparities, zhang2012vision}. However, researchers around the globe have been concentrating on these health and vision disparities and adopting various approaches to address these challenges \cite{harris2017systematic, mora2018diabetic}.

In the traditional approach to patient-physician healthcare ecosystem \cite{welcht1964appointments, wu2010overview, almalki2011health}, patients, when aware of or concerned about their medical issues, make appointments with their preferred physician(s). Upon the physician's availability of time, the patient has to travel to the physician's office. Most often this results in patients taking time off from work, arranging transportation by themselves (personal car, public transport) or by friends and families (asking for ride services). This additional resource of time, energy, and money indirectly add to the actual medical cost, which already is an existing problem in today's healthcare system \cite{torio2006national}. This hidden cost is a severe barrier for many people, especially underprivileged communities. 

To address the issue of costly healthcare for minority communities, and to provide comfortable, fast and cost-effective eye-care for diabetic people, we are proposing a collaborative telemedicine approach in community locations during regular hours and community events. On that note, although there are many software systems like Epic \cite{epic}, Cerner \cite{cerner} to provide Electronic Health Record services for clinical care, there are no established systems available to support the collaborative communication among community partners for telemedicine in community settings. Focusing that, we have designed and developed a software framework named "Mobile Teleophthalmology in Community Settings (mTOCS)". We have trained several bilingual staff from community development groups to use our system and perform eye screening events at Hispanic/Latino community events. Additionally, our system stores and provides fundus retinal images to eye-specialists, captures online diagnosis and follows up on those who were found to have retinopathy to assess best referral modality and further barriers to care.

Our research work establishes and evaluates a novel approach of providing teleophthalmology support in community settings for improving eye-health among a specific community (in our research, Latino/Hispanic). By conducting a feasibility study among 400 participants from our community screening events, we have found that this approach can significantly benefit the underprivileged people. Especially, the bilingual/native staffs using our multilingual mTOCS framework can help people to overcome the language and culture barrier of the minority communities. The pilot study conducted among the community staff has shown that the mTOCS framework can improve efficiency by saving time and providing an intuitive and usable service. Our contributions in this paper can be summarised as follows: 
\begin{itemize}[noitemsep]
    \item Proposition of a unique collaborative framework connecting the community partners with ophthalmologists for easy co-ordination
    \item Assessment of the acceptance and feasibility of organizing eye-screening events in community settings analyzed through feedback from the participants
    \item Efficacy and usability evaluation of the proposed mTCOS framework by conducting a pilot study among the community staffs who have used the framework for one year. 
\end{itemize}

In this paper, we begin with a general background, along with a discussion on problem identification and barriers. We discuss our framework with its features and uniqueness. We move onto our software deployment, usage, and data collection scenario. Our exploratory analysis of the collected data shows general population demographics and findings. We explore feedback from "participants" and evaluate the acceptance of our software framework in community settings. In the end, we discuss general take-home points along with our future steps in moving forward with our research work as well as helping general eye-health in community settings.\\ 
\section{Background}
\subsection{Diabetic Eye Disease}
Diabetic eye disease refers to the eye problems that people develop due to suffering from Diabetes for a prolonged period. Diabetic eye diseases can lead to vision impairment or even blindness \cite{niheye, kahn1974blindness}. In general, Diabetic eye disease \cite{diabetes_eye} includes: 
Diabetic Retinopathy (DR) is the most prevalent diabetic eye disease and is one of the leading factors of blindness worldwide \cite{yau2012global} as well as in American adults. It occurs due to changes in retinal blood vessels; in some cases, the blood vessels get swollen and can leak fluid. Also, abnormal growth of new blood vessels on the retinal surface can be responsible for developing DR.

\begin{wrapfigure}[12]{r}{0.55\columnwidth}
\centering
  \includegraphics[width=0.55\columnwidth]{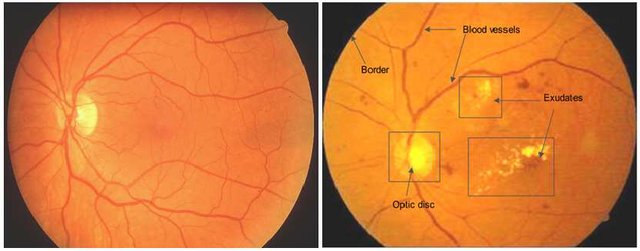}
  \caption{Fundus retinal image features in Diabetic Retinopathy subjects compared to normal retinal image (Referenced from Vimala 2014)}~\label{fig:dr}
\end{wrapfigure}
\subsubsection{Symptoms of Diabetic Retinopathy}

DR does not have any immediately noticeable symptom, e.g., pain or vision impairment, etc. during the early stages. Vision might get blurred due to Macular Edema, another eye condition which happens when DR affects macula (the part of the retina responsible for providing sharp and color vision) by swelling, due to leak of fluid. Vision is also hampered when new vessels grow on the retinal surface, causing eye bleeding. In some more complicated cases, the disease may progress without any symptom for an extended period. Due to this type of progression without any symptom, regular eye screening is vital for people with diabetes, especially for those who are suffering from diabetes for more than 15 years \cite{chaturvedi2008effect}. A comparison between a normal retina and retina with DR \cite{vimala2014economic} is presented in figure \ref{fig:dr}.

\subsubsection{Diagnosis of Diabetic Retinopathy}

While performing the annual examination of eye for diagnosis of DR, in most case, eye is dilated to enlarge the pupils so that eye-specialists can look deeper into the eyes for checking signs of DR. However, this pupil-dilation is found to be one of the barriers \cite{sopharak2008automatic} as many patients feel very uncomfortable after this dilation due to the resulting eye-sensitivity afterward. To overcome this issue of discomfort, Scanning Laser Ophthalmoscope (SLO) cameras \cite{fischer2019scanning} can be used which can capture excellent quality retinal images without this pupil-dilation. 


In summary, DR cannot be prevented; however, the risk can be greatly reduced through improving awareness about DR and emphasizing the importance of regular eye screening as well as facilitating cost-effective, accurate, comfortable solutions for the screening.

\subsection{Community Telemedicine}

To overcome the existing vision health disparities \cite{zhang2012vision} among socioeconomically disadvantaged groups, a paradigm shift through innovative interventions is necessary. Community telemedicine refers to a collaborative approach through creating a multi-sector research team representing distinct areas of expertise where each individual serves a specific role in building a strong and effective partnership. This approach can address health disparities and improve eye health by focusing on individual communities\cite{harris2017systematic}. Hispanic/Latinos are identified as the high-risk population for diabetes\cite{mora2018diabetic} and diabetic eye diseases. They also face few additional constraints resulting from socioeconomic, cultural challenges, and a lack in the number of Spanish-speaking eye-specialists\cite{derose2000limited,ortega2007health}. Compared to English-speaking, non-Latino counterparts, non-English-speaking Latinos reports 22\% fewer visits to physicians in the state of Wisconsin, USA \cite{wisconsin_latino}. Therefore, to address the health, particularly, vision health disparities, a collaborative telemedicine approach can facilitate the improvement of eye-health among some specifically disadvantaged communities by providing special attention to their custom barriers. 

\subsection{mHealth Tools and Limitations}

Availability of internet and IT tools, access to large-scale datasets \cite{johnson2016mimic}, improvement in diagnosis accuracy by novel machine learning models \cite{chen2017disease}, artificially intelligent systems \cite{jiang2017artificial}, and finally, the growth of digital technology, in general, has led recent research to be focused on development of mobile health (mHealth) tools to aid in healthcare systems. Specialized mHealth tools \cite{bolster2016diabetic} have been built for the treatment of DR; however, they have been focusing on traditional clinical solutions with minimal focus on human-computer interactions. Access to affordable healthcare for all is a large crisis \cite{kaiser_uninsured}, and existing works fails to address issues that possibly could mitigate it. One of the existing mHealth solutions, EyePACS \cite{cuadros2009eyepacs}, is also focused on improving retinal care for diabetic patients, and have been working actively with clinicians to provide a fast diagnosis. However, our proposed framework focuses more on the active roles of community partners as a facilitator of connection and communication with specific communities. In addition to that, considering the barrier of healthcare insurance cost \cite{hadley1991comparison}, our framework also provides such care outside the hospital in a community event, where more people with no health insurance can come and do eye-health checkups. Finally, existing Electronic Health Record (EHR) systems, such as Epic \cite{johnson2016comprehensive}, has aided in efficient healthcare support in general; however, this is more of a general tool and is not fully customized to support focused community groups.  

\subsection{Our Contribution}

There exists a minimal but emerging, HCI literature in recent years that considers the use of technology to mitigate the epidemic effects of diabetes among people. Several research groups are working on the self-management of diabetes \cite{raj2019my, desai2019personal}, as well as some other groups have proposed interactive frameworks to provide better education about diabetes \cite{katz2018designing, desai2018pictures, kyfonidis2019making, katz2018data}. However, to bring down the percentage of vision loss resulting from diabetes, we also need to focus on educating people about the importance of annually recommended eye-screening. Our unique contribution aligns with this motivation. To summarise, our exploratory background study shows us the following barriers to affordable and regular eye-screening: 1) financial barrier: \textit{lack of health insurance}, 2) communication barrier: \textit{disconnection due to language}, 3) motivation barrier: \textit{lack of motivation to make appointments with doctors, discomfort in clinical settings}, and 4) knowledge barrier: \textit{lack of health knowledge in minority communities}. Our proposed software framework, mTOCS, is designed to address these findings and is built as a community-focused mHealth tool, providing telemedicine support to diabetic patients with eye-health issues. 
\section{System Design}

Our mTOCS software framework has been designed with a focus on simplicity, efficiency and efficacy on promoting the collaboration between multiple moving agents (screeners, graders, participants, etc.). This section describes the design methodology of the software framework along with the different roles with their actions and access levels, and the features that makes it beneficial for conducting eye-screening events in community settings. 

\subsection{User Roles, Access Levels and Functionalities}

The unique strength of our mTOCS software framework is the availability and support in community settings. One of the goals of "Tele-eye Health Project" is not to bring people to a clinical setting for a service; rather, we brought service to the community people in their comfort zone. We spread the news of free eye-screening events offline through word-of-mouth, fliers, social events as well as online by posting on social media and advertisements with the support from the City Health Department. The screening events were held in community centers, health fares, food panties, schools, Health Department clinics, etc. Our focus is to cover the most convenient community locations. Each screening event continues for 3 or 4 hours, covering 18-20 participants in each event. 
\begin{wrapfigure}[13]{l}{0.5\textwidth}
    \centering
    \includegraphics[width=0.5\textwidth]{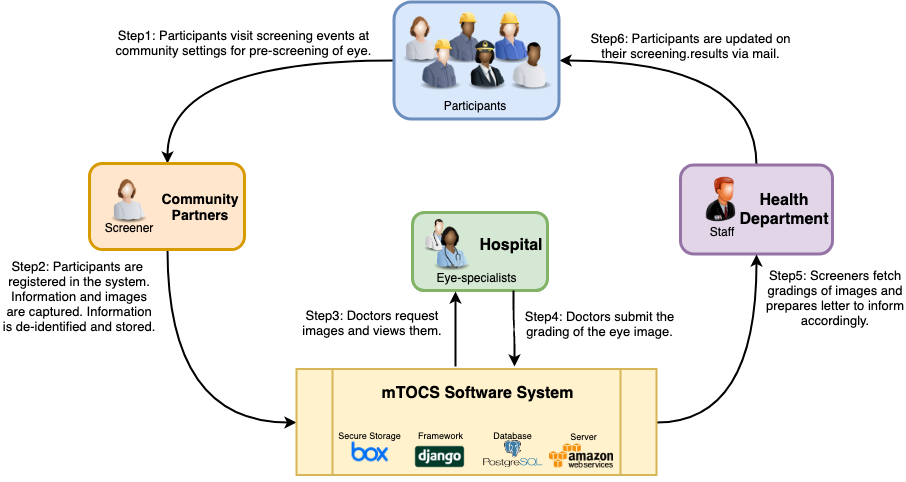}
    \caption{Work flow in mTOCS}~\label{fig:workflow}
\end{wrapfigure}

In this project, \textbf{`participants'} refer to general people seeking retinal screening at a community event, \textbf{`screeners'} refer to staffs from community partners (e.g., United Community Center, Milwaukee, USA) using our software and managing the complete screening process (data entry, DR image collection and storage, etc.), \textbf{`graders'} refer to specialized physicians (e.g. ophthalmologists) who diagnose and grade the retinal images, and \textbf{`staffs'} refer to general support staffs (e.g. Milwaukee Health Department) who receive the grading and sends out relevant letters with diagnosis and recommendations to the participants. Difference between screeners and staffs is that, screeners actively work as community support whereas staffs work behind the curtain and maintain official tasks as needed. We present the complete workflow in figure \ref{fig:workflow}.

A general scenario of an eye-screening goes as follows: participants were greeted at the reception and added to a queue where they can wait for their turn to come. The event locations had arrangements of adequate seating space in a comfortable waiting area. When a participant's turn comes up, they were called and seated in a separate area. The screeners greeted them and started to demonstrate our general goals and steps. Their demographic information is collected and stored using our mTOCS software, followed by a survey of eye-health. The survey and discussion were continued in the preferred language of the participant, in which he/she was most comfortable and fluent. After that, the participants ware requested to move to a separate dark room with SLO fundus camera set up.  Images were taken using the SLO fundus camera where screeners also entered the survey id in the camera software for future reference of the images. The SLO fundus camera used in our screening events captures non-mydriatic fundus images so that participants visibility is not obstructed for the next few hours. After finishing up with capturing of the fundus images, the participants are informed about the time when he/she should expect screening results along with the future updates of screening events.

Depending on the responsibilities performed by the community partners and eye-specialists involved in 'Tele-eye Health Project,' the mTOCS framework has four different access roles with distinguishing features and support: 

\begin{enumerate}
    \item \textbf{Screener Portal \textit{(Only accessed by screeners)}:} Participant registration (with demographic information), Language preference, Survey completion, DR image capture and storage
    \item \textbf{Grader portal \textit{(Only accessed by graders)}:} View ungraded image, Grade images, Change/update graded images
    \item \textbf{Report Distribution Portal \textit{(Only accessed by staffs)}:} View retinal image grading, Easy printout of preformatted letters with image grading, Future communication, Request to follow-up
    \item \textbf{Data management Portal \textit{(Only accessed by system admin)}:} Data export/ backup procedure, Data analysis
\end{enumerate}

\begin{figure}[ht]
\centering

\begin{subfigure}{0.48\textwidth}
\centering
\includegraphics[width=\linewidth]{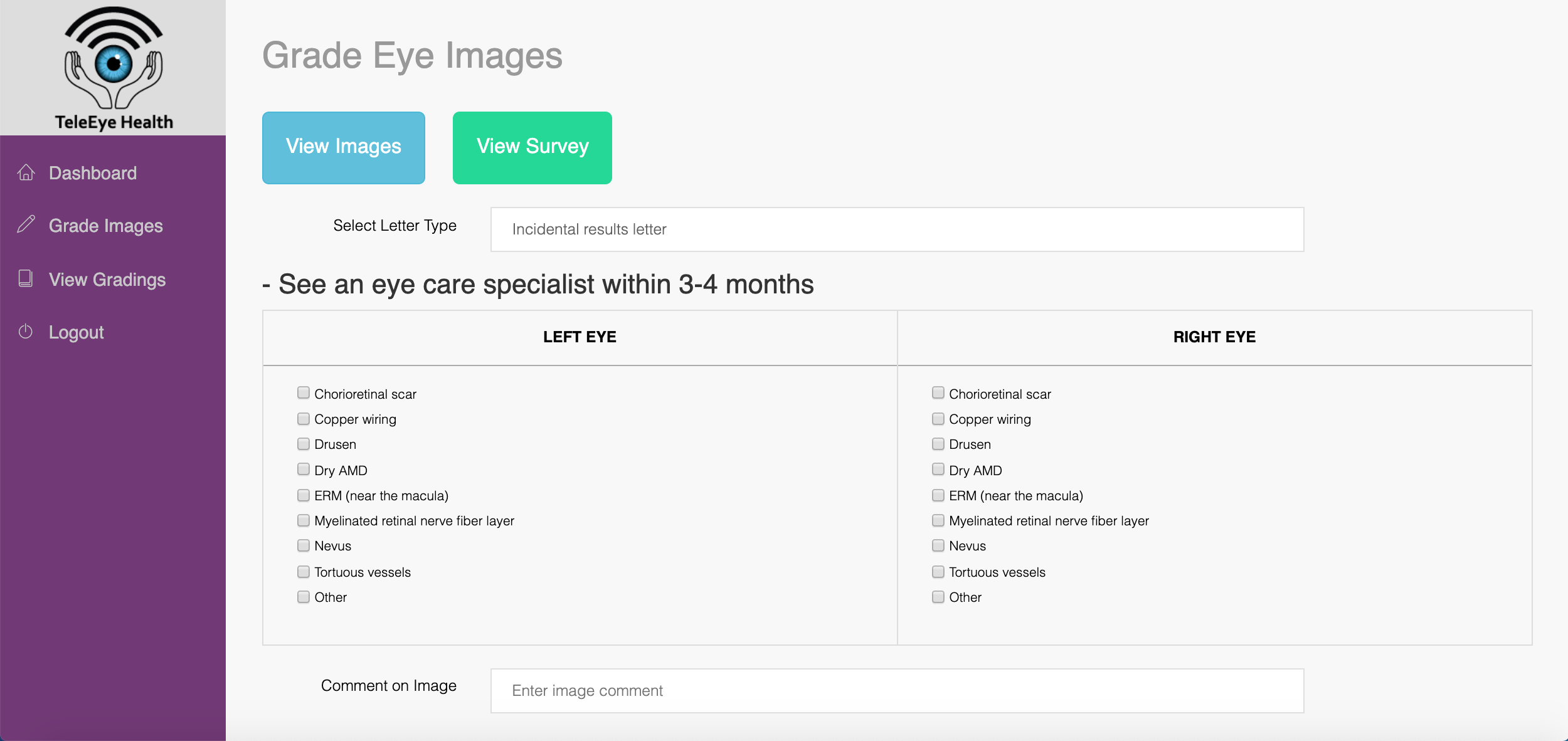} 
\caption{Grader portal: List of features of Diabetic Retinopathy shown, depending on letter type selected}
\label{fig:grader1}
\end{subfigure}
\begin{subfigure}{0.48\textwidth}
\centering
\includegraphics[width=\linewidth]{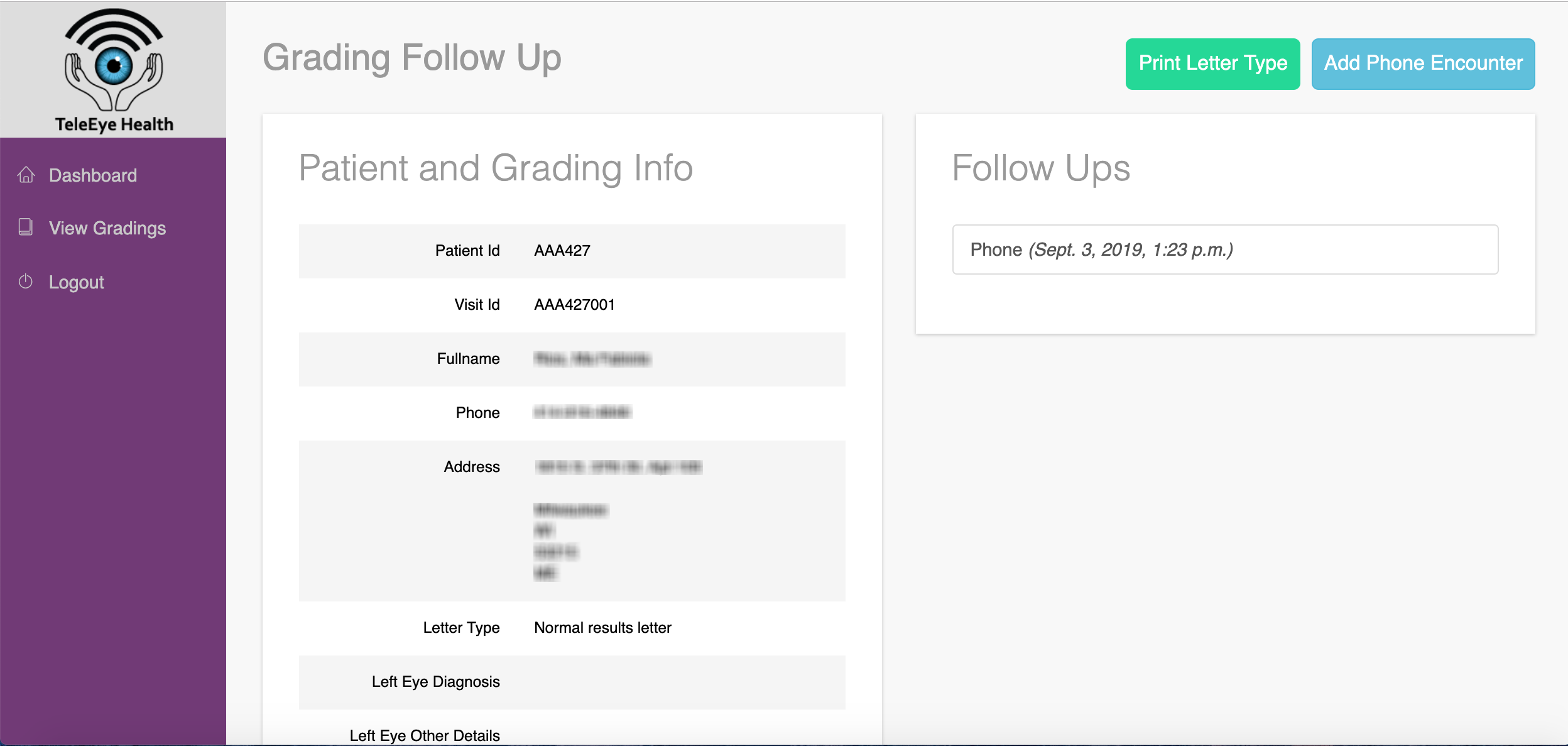}
\caption{Staff portal: Grading follow-up scenario for a certain participant}
\label{fig:mhd}
\end{subfigure}

\caption{Screenshots from proposed mTOCS framework}
\label{fig:software}
\end{figure}

\subsection{Design Motivation}

Before designing and development of our software system, "Tele-eye Health Project" officially started to arrange the screening events in different community locations of Milwaukee, Wisconsin, USA during June 2017. For the first six months, participant's information and questionnaires were collected through traditional paper forms. During this time, the community partners and eye-specialists faced multiple barriers which lead to the development of our framework 'mTOCS'. 

\subsubsection{Redundant Time Consumption}

From the screener's (staffs of community partners) perspective, the use of paper forms caused significant inconvenience in terms of time and efficiency.  After conducting the screening events, the screeners had to enter all the information into a Microsoft Excel sheet at the day end. Then, these sheets were uploaded to the private account of secured storage in "Box.com" \cite{box} and all the paper forms were shredded for the privacy concern. In this way, the screeners had a vast workload even after the end of the tiring screening events; however, this data-entry part had no significant contribution to the overall workflow.

From the grader's i.e eye specialist's perspective, it was also very inconvenient to open an excel sheet to look into participant's history and then open the images from the cloud storage by matching the participant's de-identifier manually and then, enter the grading results and comments into excel sheet and finally, updating that excel sheet to the cloud again to make it available for the community partners so that they can send those results back to the participants. It is also noteworthy that managing multiple versions of excel sheets across various users is a very manual and cumbersome process.

While collecting the requirements of the portal of community partners, staffs of health department mentioned that when they send the results back to the participants, they need to find out the address of the participants from the excel sheets containing participants' information which is very inefficient, especially since our dataset is growing cumulatively. Moreover, since there are eight different types of letters that are sent to participants depending on their eye conditions and all of them can be both in English and Spanish, so it takes a lot of clerical work to find out the appropriate letter depending on the grading and language mentioned in the excel sheets containing results graded by eye-specialists and then printing that letter, and finally put the address found from another excel sheet containing participant's information. 

\subsubsection{Error occurrence}

Staffs of our community partners mentioned that while entering and retrieving to and from Excel sheets, it often leads them to neighboring rows by mistake as we have around twenty different columns for each participant. This type of error is acutely endangering here, since it is related to a person's health, and getting a wrong letter mentioning eye condition result might cause severe consequences. Thus, the staff addressed the importance of having a portal where they can view a particular participant's address and necessary information by searching using relevant keyword like, name, de-identifier, phone number, email, etc. They also mentioned that it would greatly reduce their error rate if they could print that particular letter from mTOCS where the letter is already selected by the system depending on the grading result and language. 

\subsection{Action Research Methodology}

Action Research (AR) Methodology \cite{olson2014ways} has been adopted for the design of mTOCS framework to support the collaborative research work of Tele-eye Health project. mTOCS has been developed with multiple iterations by knowing the needs and requirements of our Eye-doctors and Community partners. As described in the following sections, mTOCS has four different access roles. Our research team has collected requirements from the direct users of respective portals, analyzed those and finally, designed the features to accommodate best to their needs at field level. All the portals have gone through multiple iterations of feedback, and reflections have been made on that feedback. One relelvant example of application of AR for mTOCS is that, the Data Management portal which was not included in our initial plan. When our community partners started the screening event at different community locations using our mTOCS framework, they often asked the system administrators to send all data to them so that they can do statistical analysis to understand the impact of different locations in the community better. Since our collaborators work from various remote locations, this transfer of data had to be completed through any of the email service or file sharing service, that prioritizes privacy concern. Thus, our research team integrated another portal where an authorized user can directly download the updated data from the database and can perform analysis. This portal has been designed in a minimalist way so that it is not overused and are accessed by the authorized users only when necessary. 

\subsection{Efficiency Catalysts}

mTOCS has been designed and implemented with a concentrated focus on improving the efficiency of the involved users and providing users with comfort and reducing clerical burdens. Some of the worth-mentioning efficiency catalysts \textit{(features that improved efficiency)} are discussed in this subsection.

\begin{itemize}
    \item \textbf{Questionnaire Auto Show/Hide:} In the questionnaire, there are some subsections which depend on the response of another previous question. For example, if a participant responds a 'Yes' to the question about having diabetes, only then, the question regarding the type, duration, etc. about diabetes are asked. While designing mTOCS, the dependent subsections are automatically shown or hidden, depending on the response. The screeners do not need to skip the questions by themselves; rather, the design of mTOCS supports them for surveying in a compact and organized manner.
    \item \textbf{Easy Search of Returning Participants:} The screeners do not need to take all the demographic information for a returning participant that already exist in the system. To accomplish that, the mTOCS software supports the "Search" feature so that the screeners can easily search a participant when they visit again for an eye exam. A participant can be searched in the system using multiple keywords. The software supports search with auto-generated participant ID, first or last name, date of birth, or phone number of the participant.
\end{itemize}

\subsection{Strengths of the System}

We summarize the unique features of our mTOCS software system, that facilitates better coordination among collaborators and makes mTOCS a unique and robust framework for fast and efficient collaboration.

    \begin{itemize}
        \item \textbf{Language Support:} The questionnaire of the survey and the grading result letter are both in English and Spanish language so that the staffs can get conduct the survey more comfortably using our framework.
        \item \textbf{Scalability:} The back-end design of the software is modular and scalable. There is a hierarchy among the modules, and a full workflow, as shown in figure \ref{fig:workflow}, is under an organization. If any new organization gets involved in our project, we can easily integrate them into our system as an independent part by providing custom secured credentials to the new organization. Overall, modular design structure makes the addition of any new organization or language highly fast and easy in our system.
        \item \textbf{Security:} The system uses PostgreSQL as a database, and the complete system runs into an Amazon Web Services (AWS) cloud. The system can only be securely accessed through HTTPS connection. The software is also behind the Amazon firewall, which implies every connection other than HTTPS will be blocked, and the database is inaccessible from the outside network. The collected retinal images are stored into the Box.com server, a third-party secure cloud storage provider. The system communicates with the Box server using Box APIs, which are also secured by HTTPS. As the system keeps a record of demographic information of participants, we comply with HIPAA guidelines for proper storage and access to database.
        \item \textbf{Privacy:} The permission to access different portals of the mTOCS system is completely different and depends on the various roles. User functionality is limited through user access role. Moreover, the eye-specialists are also unable to view the private information of the participant while grading the retinal images. The auto-generated ID that used to refer a participant is also de-identifiable.
    \end{itemize}

\subsection{Deployment of mTOCS Software}

\begin{wrapfigure}[12]{r}{0.45\columnwidth}
\centering
  \includegraphics[width=0.45\columnwidth]{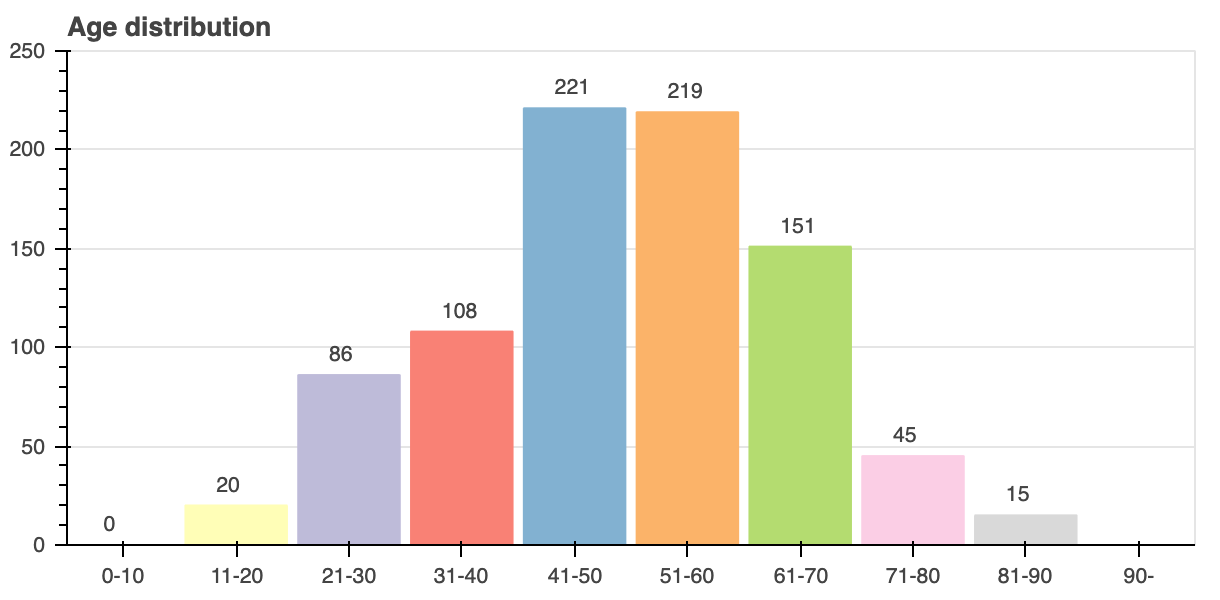}
  \caption{Age distribution of participants}~\label{fig:age}
\end{wrapfigure}

After our software research and development cycle was completed, we deployed a test version of the system for the screeners to test it in simulated scenarios. The screeners were asked to perform the following functionalities: 1) test run the software to train the screeners, 2) find out features to add/improve upon for increasing the intuitiveness of the system, and, 3) check for any potential bugs. We iterated the process multiple times to make sure of covering every possible screening scenario in our software. We also communicated and discussed the newly generated issues with eye-specialists to keep them updated on the changes made on the software. After the development and testing phase, we reset the complete database and made it open for regular private-access use.

Since the start of the paper-based screening events, our system has screened 1402 participants over multiple screening events at multiple sites at Milwaukee, Wisconsin, USA. Our mTOCS system was deployed in January 2018, through which we have screened 845 individuals through a number of screening events. A description of dataset follows in the next segment.
\section{Exploratory Analysis of Database}

Since January 2018, our mTOCS software system has been used in 53 community screening events over different places in Milwaukee, Wisconsin, USA. 6 screeners have been registered to have access in our system, and have been using our software to capture screening information of 1402 participants till date.

\subsubsection{Description of Participant Demographic Dataset}

\begin{wraptable}[34]{r}{0.5\columnwidth}
  \centering
  \begin{tabular}{ l r }
    \hline
    \textbf{Age} & \textit{mean (range)}\\
    & 49.91 (15-89)\\
    \hline
    \textbf{Gender} & \textit{count (\%)} \\ 
    Female & 530 (61.27\%) \\
    Male & 335 (38.73\%) \\
    \hline
    \textbf{Ethnicity} & \textit{count (\%)} \\
    Hispanic or Latino & 317 (36.65\%) \\
    Black or African American & 245 (28.32\%) \\
    White & 117 (13.53\%) \\
    Asian or Pacific Islander & 95 (10.98\%) \\
    Native American & 76 (8.79\%) \\
    \hline
    \textbf{Country} & \textit{count (\%)} \\
    USA & 483 (55.84\%) \\
    Mexico & 253 (29.25\%) \\
    Myanmar (Burma) & 24 (2.77\%) \\
    Laos & 21 (2.43\%) \\
    Puerto Rico & 8 (0.92\%) \\
    South America & 4 (0.46\%) \\
    Other Central America & 14 (1.62\%)\\
    Other Carribean & 1 (0.62\%)\\
    Other & 15 (1.73\%) \\
    \hline
    \textbf{Insurance} & \textit{count (\%)} \\
    No insurance & 341 (39.42\%)\\
    Private insurance & 224 (25.90\%)\\
    Medicare (65 years or older) & 58 (6.71\%)\\
    Medicaid (T-19/ Forward card/ low-income) & 146 (16.88\%)\\
    Medicare and Medicaid & 44 (5.09\%)\\
    Other insurance (e.g. VA, TRICARE) & 52 (6.01\%)\\
    \hline
    \textbf{Language} & \textit{count (\%)} \\
    English & 502 (58.04\%)\\
    Spanish & 265 (30.64\%)\\
    Both & 13 (1.50\%)\\
    Other & 85 (9.83\%)\\
    \hline
  \end{tabular}
  \caption{Demographic distribution of participants}
  ~\label{tab:demographic}
\end{wraptable}
Since our focus is to help people with long-time diabetes and higher aged people has more complications from diabetes and more likely to have eye issues, our participant database mostly consists of people with higher age. The age distribution is more skewed towards higher age, which is shown in figure \ref{fig:age}. The average age is about 50 years old, ranging from 15 years to 89 years. About 61.27\% of the participants were women. More than half of the participant population were Hispanic/Latino (36.67\%) or African American (28.32\%). Although more than half of our dataset consists of participants who are the citizen of the USA (55.84\%), participants from Mexico (29.25\%) was second in the majority. Around 40\% of participants had no health insurance coverage, and about 30\% of participants were only Spanish-speaking. A more specific overview of the participant characteristics is presented in table \ref{tab:demographic}.

\subsubsection{Overview of Exploratory Findings}

Our exploratory analysis draws some crucial conclusions from our demographic dataset.

\begin{itemize}
    \item Though we have worked with community partners from Hispanic/Latino community, it was open for people from any community. When we looked into the overview of our dataset, we have found that African American population has the second largest percentage among our total participants. This finding directs us to the hypothesis that African American community should also be focused next while expanding this project.
    \item The percentage of people within the age-range of 40 to 60 is the highest. Also, people already having diabetes occupies a majority portion of our population which hypothesizes that we have been able to reach our target population. 
\end{itemize}

\section{Framework Evaluation Study}

Our tele-eye health project, mTOCS has multiple dimensions of user interactions; on one end, it has to perform fluently as part of a conversation piece in an interview setting with the mass people, on the other hand, it serves as an intuitive tool to be used by the screeners without minimal disruption. To capture the strengths of these dimensions and to evaluate the impact of our proposed software framework, we developed and conducted a participant satisfaction survey, addressing the comfort and convenience of communities in a tele-eye health project. 

\subsection{Participant Satisfaction Survey}

In the participant satisfaction survey, through collecting general participant feedback, we wanted to explore whether and if the participants could accept eye-health screening conducted publicly. Our goal was 1) to assess the acceptance level and attitudes of participants regarding teleophthalmology in community health settings in Milwaukee, 2) to analyze the strengths and weaknesses of the eye-screening events through comparing likes and dislikes among participants.

\subsubsection{Measures}

The survey contained a total of eleven(11) questions. Eight(8) of them are 5-point Likert scale questions, addressing various aspects of the community setting eye-screening which includes comfort, location, privacy, involvement, dissemination, connection, language, and general acceptance. Options of responses ranged from 1 (strongly disagree) to 5 (strongly agree). Details of each statement and the corresponding theme are presented in table \ref{tab:questions}. The other three(3) are open questions about what the participants liked about the screening, disliked about it and whether they would discuss it (or recommend it) to their friends and families. The survey does not directly focus only on the software, instead discusses issues that connect the framework as a whole. This involves the whole ecosystem consisting of location, timing, use of camera hardware-software combination, and screening procedure in general. 

The participants were asked to fill up the complete satisfaction survey after they completed their free eye-screening. Although self-reported survey responses are not always the best representation of the entire scenario, however, they can convey critical information and valuable feedback from a different perspective, and points towards a general acceptance of the targeted community.

\begin{table}[h]
  \centering
  \begin{tabular}{| l | l |}
    \hline
    \textbf{Statement} & \textbf{Theme}\\
    \hline
    \textit{I was comfortable during the telemedicine} & Comfort\\
    \textit{session} & \\
    \hline
    \textit{It was convenient to get eye screening at} & Location\\
    \textit{this location} & \\
    \hline
    \textit{I was worried about my privacy using this} & Privacy\\
    \textit{type of eye screening} &\\
    \hline
    \textit{This type of screening helped me get more} & Involvement \\
    \textit{involved with my health} &\\
    \hline
    \textit{I would recommend this type of screening} & Dissemination\\
    \textit{to others} &\\
    \hline
    \textit{I liked seeing the image of my retina} & Connection\\
    &\\
    \hline
    \textit{It helped having bilingual/ Spanish staff} & Language\\
    \textit{do the screening} &\\
    \hline
    \textit{I would use this type of screening again} & General\\
    & acceptance\\
    \hline
  \end{tabular}
  \caption{Satisfaction survey questionnaire}
  ~\label{tab:questions}
\end{table}

\subsubsection{Methods}

A total of 400 satisfaction survey were given to participants after completing the retinal screening process in either English or Spanish. Excluding the incomplete or no responses, we collected a total of 378 complete satisfaction survey responses filled up by participants. The surveys were provided to the participants at the end of the respective eye-screening and accumulated through multiple eye-screening events sessions. On average, the participants were comfortable throughout the screening process. 

\subsubsection{Study Results}

The response rate for our satisfaction survey were 94.5\% (378 out of 400). We had a total of 225 (59.52\%) participants preferring language English for communication, 103 (27.25\%) participants preferring language Spanish for communication, and 50 (13.23\%) participants with no preference. The participants with no reference indicate either being comfortable in both languages, having no preference over language or unwilling to submit a response to this specific question. 

\begin{figure*}[h]
\centering
  \includegraphics[width=\columnwidth]{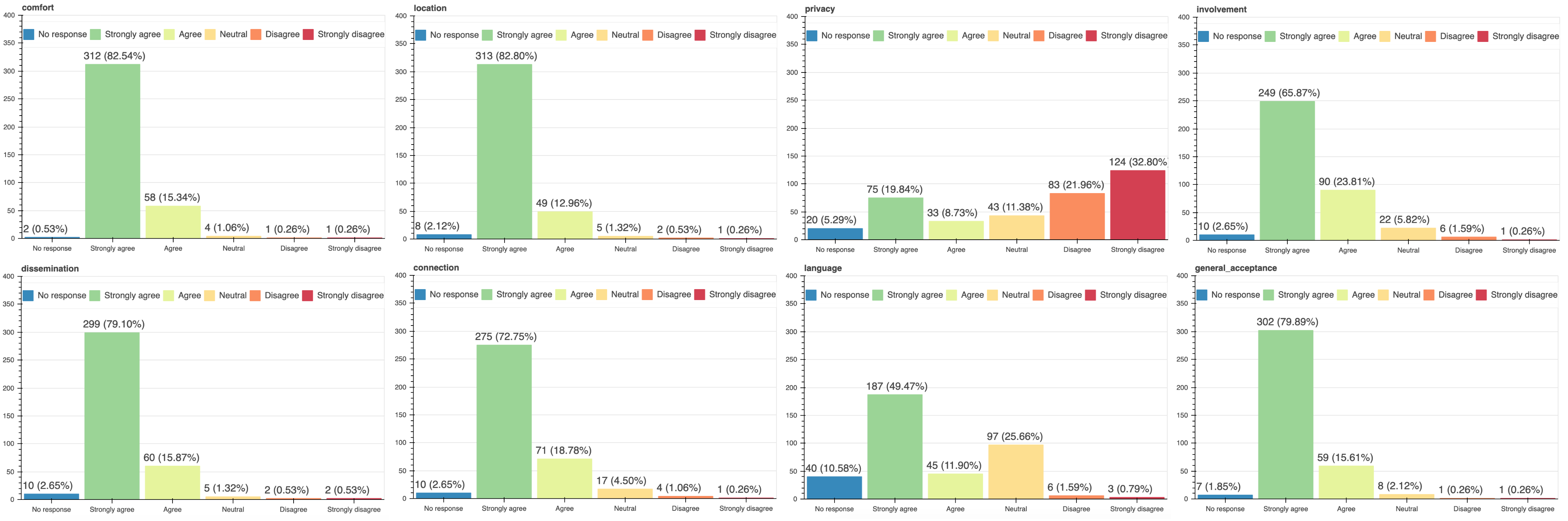}
  \caption{Response representation of satisfaction survey questionnaire}~\label{fig:barplots}
\end{figure*}

In general, the responses from the participants were very positive. The responses for individual eight Likert-scale questions are represented in graphs in the figure \ref{fig:barplots}. Calculating based on a scoring of 1 for a response of "strongly disagree" to 5 for "strongly agree," a mean (average) score of more than 3 signifies a general positive response towards the specific question. The number of responses and their mean, standard deviation for individual questions is described in table \ref{tab:meanmedian}.

\begin{table}[h]
  \centering
  \begin{tabular}{| l | c | c | c |}
    \hline
    \textbf{Theme} & \textbf{Count} & \textbf{Mean} & \textbf{Standard Deviation}\\
    \hline
    Comfort & 376 & 4.81 & 0.48 \\
    \hline
    Location & 370 & 4.81 & 0.49 \\
    \hline
    Privacy & 358 & 2.58 & 1.54 \\
    \hline
    Involvement & 368 & 4.58 & 0.70 \\
    \hline
    Dissemination & 368 & 4.77 & 0.55 \\
    \hline
    Connection & 368 & 4.67 & 0.64 \\
    \hline
    Language & 338 & 4.20 & 0.98 \\
    \hline
    General acceptance & 371 & 4.78 & 0.51 \\
    \hline
  \end{tabular}
  \caption{Quantitative representation of responses for satisfaction survey}
  ~\label{tab:meanmedian}
\end{table}

Based on the average mean of a score higher than 4, we find that in general, the community eye-screening event satisfies most of our participants through different directions. The question about privacy and language does not have a high positive response, as seen in other questions. Unlike other questions, the question regarding privacy (\textit{"I was worried about my privacy using this"}) is negatively constructed. Because of this, a low score (2.58, which is less than 3) actually signifies that the participants are not much worried about their privacy. However, we also find a difference of opinion on that in between participants.

\begin{wrapfigure}[13]{r}{0.5\textwidth}
\centering
  \includegraphics[width=0.5\textwidth]{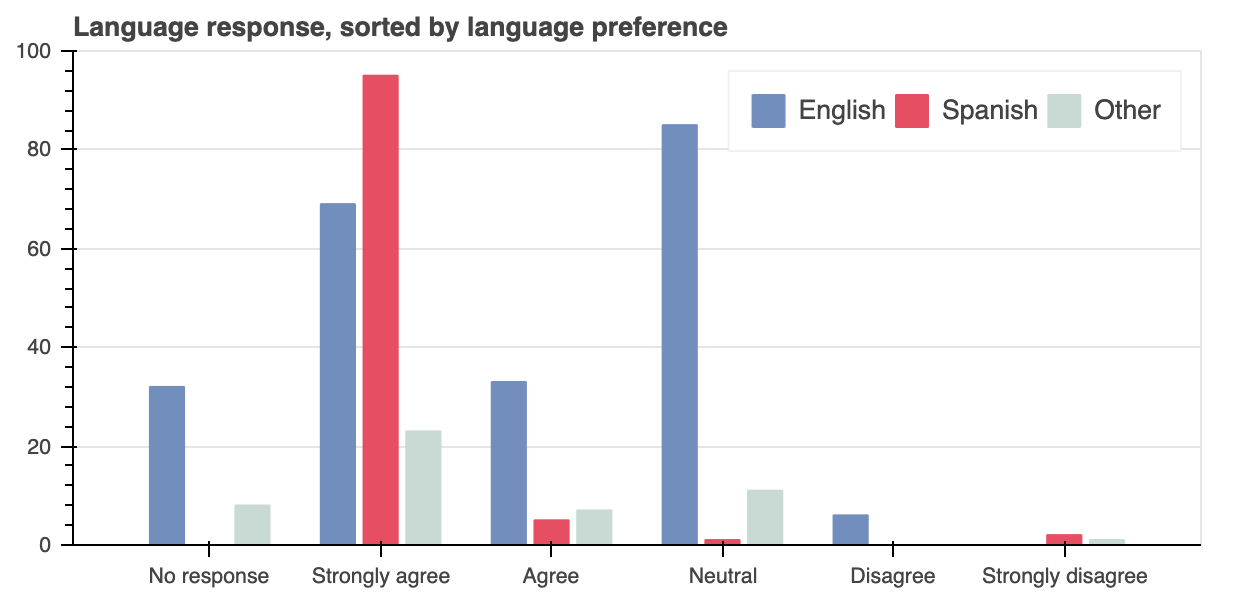}
  \caption{Response for question theme "Language", stratified by language preference}~\label{fig:language}
\end{wrapfigure}

An interesting finding is that, for the question about language (\textit{"It helped having bilingual/ Spanish staff"}), we find two peak answers, "strongly agree" and "neutral." Further exploration into these responses helped us see that language-preference of a participant plays an active role in the response of this question. If we break down the responses of participants for this question based on their preferred language for communication, as shown in figure \ref{fig:language}, we find that, most participant who preferred Spanish for communication agrees with the statement (\textit{count of "strongly agree" = 95}). Compared to that, participant group for preferred communication language English and others is mostly indifferent ("neutral") to it. The general response to this question is confounded by their preferred language of communication. This indicates that, although our software helps to connect with Spanish-speaking people, most people who are unaware of language barriers are indifferent about these additional options.

In our satisfaction survey, additional to the eight(8) theme-specific questions, we had three(3) other free-text questions. The questions asked the participants general feedback about these following topics:

\begin{itemize}
    \item \textbf{Likes:} \textit{Did you like this type of screening? What will you tell your family and friends about this type of screening?}
    \item \textbf{Dislikes:} \textit{What did you like the least? How could we improve this process?}
    \item \textbf{Referrals:} \textit{Will you talk to your family and friends about this screening? Any other comments?}
\end{itemize}

In general, most of the participants expressed their satisfaction for the process, and few shared what could be improved. 



In expressing positive feedback, we received comments like this:

\textit{"si me gusta el programa y le diria a mis amigos y familia que se hagan este examen. Lo considero muy importante sobre todo para las personas prediabeticas." (Yes, I like the program and I would recommend this exam to my friends and family. I think it is very important especially for the people that are pre-diabetic)}

However, when expressing their dislikes, we primarily received feedback on three types of discomforts.

\begin{itemize}
    \item \textbf{ Camera technology:} For some people, the SLO fundus camera flash was perceived as "too bright." The SLO fundus camera captures images of the retina in a non-mydriatic way (without dilating the pupil), however to do so, it has to use a bright flash for a very short timespan. Example: \textit{"the light is too bright", "la luz es un poco incomoda (The light is a little bit uncomfortable)"}
    \item \textbf{ Equipment limitation:} The cost of a regular SLO fundus is very high, because of which we had to use a single camera for a whole event. Many people had to wait for a longer period, compared to traditional doctor visit waiting time. Example: \textit{"the wait time due to numerous people, but I was okay with the air conditioning"}
    \item \textbf{ Response time:} Response with the analyzed result took a bit longer, compared to traditional doctor check-up. This is primarily because the eye-specialist grades the images at his/her convenience and free time, so immediate response was not feasible. Example: \textit{"waiting on results"}
\end{itemize}

\subsection{Screener Software Assessment}

Compared to the number of participants, we have limited number of human-resources who acts as screeners using the mTOCS software in the eye-screening events. To capture their viewpoint, we conducted a separate pilot study focusing on general software assessment, focusing on four different aspects of our software system. The four focus points for evaluation of our system from the viewpoints of screeners are (1) feasibility, (2) usability, (3) viability, and, (4) satisfaction. This feedback added value to our system structure and project continuation, since the screeners have went through iterations of the mTOCS software and are the perfect candidate to be representing the other dimension of the software.

\subsubsection{Measures}

The survey was designed starting with a general introductory section. It consisted of demographic questions regarding the screeners participating the survey. This section collected information on age group (18-24, 25-34, 35-44, 45-54, and 55+) and the survey-participating screeners' technical proficiency in using new softwares, in a Likert Scale question. Age and technical proficiency are generally found to be correlated \cite{knowles2018wisdom, gell2015patterns} and are both potential indicators of successful field usage of our software system.


Evaluation protocols for similar software systems and HCI toolkits have been extensively researched recently \cite{ledo2018evaluation}, and our survey was designed closely following these findings. Ledo et al. \cite{ledo2018evaluation} have broadly summarised evaluation strategies into questions regarding four categories, (1) demonstration, (2) usage, (3) performance and (4) heuristics. Since our research questions do not focus on the heuristics, this survey was designed to address demonstration, usage and performance of our software system in real-world scenario.



\subsubsection{Methods}

Although our software system was being used to serve more than 1400 people from community, access to it is limited to a handful of trained screeners. We conducted our screener software assessment survey to only those who actively used in the field and collected data using the software. A total of five (5) complete responses were collected from five individual screeners. The screeners gave consent to collect their responses and in general it took 15-20 minutes to complete the complete set of questionnaire.

\subsubsection{Study Results}

Since it was a pilot study consisting of five (5) screeners only, the study result was pretty much straightforward and overall positive. We addressed the actual response they agreed to within parentheses in the following parts.

The screeners were of diverse age group (1 in age 18-24 group, 2 in age 25-34 group, 1 in age 35-44 group, and 1 in age 45-54 group). Two of them had strong technical proficiency \textit{(I figure out all functionalities with little effort all by myself)}, two had fair technical proficiency \textit{(It takes effort, but I get all the features by myself)} and one had average technical proficiency \textit{(I can use basic features, but often need training for advanced features)}.

Under feasibility, 80\% mentioned the software saved a lot of their time. On average, completing the survey with paper took 15.5 minutes whereas the same using software took only 6.5 minutes, which is less than half of paper survey. 80\% rated the software highly efficient compared to the paper form.

Under usability, 60\% of screener survey participants marked our mTOCS software system as highly usable \textit{(I could use all functionalities with ease)} as well as highly intuitive \textit{(I could figure out all functionalities very easily without any help)}. Even though, all of them needed a little training using the software \textit{(I needed training to figure out the basic features)}.

Under viability and satisfaction, 60\% mentioned our software still had a little bit of errors \textit{(The software runs smoothly most of the time, with a few hiccups)}, and the other 40\% mentioned facing lesser number of errors \textit{(We find once a few error here and there)}. 80\% were also highly satisfied with the mTOCS software, based on its user interface, features and in general, performance, and 20\% were somewhat satisfied.

Other than Likert-scale responses, we additionally collected general feedback and suggestions, which consisted of constructive criticisms. 
\begin{itemize}
    \item While discussing about feedback on software features, they mentioned our framework saving time while they focused on other tasks in the screening process \textit{(``It is very intuitive and allows for screeners to spend more time in other aspects of the screening process such as educating the participants and taking the retinal images'')}.
    \item On enhancing viability of our software system, screeners requested a generic dashboard with an overview of general data collected \textit{(``It could be help if we had a dashboard where we could see some of the analytics/data of the software, how many need to be read, how many need to be called back, etc. ...'')}. This is kept as part of future work, since we are investigating the most important facts, which, when displayed on dashboard, would aid in screening process as well as motivate them in their community aid work.
    \item On dislikes about our software, one screener mentioned disruption due to loss of internet connectivity \textit{(``During events if we lose access to internet then we need to resort back to the paper'')}. This is an important issue which goes beyond our software support, and we plan to counter it by providing better internet connectivity, even in the most rural and disconnect areas.
\end{itemize}
\section{Discussion and Future Work}

This paper explores a collaborative telemedicine approach in community settings to facilitate better eye-screening focusing on Hispanic/Latino Community. From June 2017 to June 2019, 1402 people have been screened through our tele-eye health project.  We have explored a unique partnership structure that connected the clinical expertise of an eye-specialist with the community-engaged research expertise of nonclinical researchers. We get an overview of the distinctions between clinical and community-engaged research, which focuses on the importance of community-engaged intervention to better-serve a community. Some initial insights from the perspective of our Medical and Community partners have been discussed in our previous work \cite{denomie2019lessons}. Here, we discuss a few other ideas and insights gathered from the project through some analysis, and also, we present the limitations and future works planned in this project direction.

\subsection{Technology to Improve Coordination and Efficiency}

Community-based telemedicine has been identified to be a feasible mechanism for conducting diabetic eye screening and for working with diverse communities. The computer-aided screening system has significantly improved the coordination among community partners and also has improved the efficiency of the community staffs. Especially from the feedback of our partnered eye-specialists, we realize that our software framework mTOCS has helped them in better time-management in spite of their hectic schedule, since they can do the grading of the retinal images in their preferred time. mTOCS has also reduced the clerical burdens of the community staffs by getting them rid of manual data entry and data retrieval steps. This has contributed to having more enthusiast and empathetic behavior with the eye-screening participants. 

\subsection{Engaging the Community Partners for Improved User Experience}

While designing the mTOCS software framework, we have highly focused on accommodating the comfort and ease of use for the community partners. To accomplish that, we have arranged multiple collaborative sessions to determine the structure and workflow of mTOCS. Instead of building a framework first and then, training the community partner staffs to use our framework, we have adopted a different principle of designing the framework following the community partner staff's preferred workflow. This principle resulted in having fewer iterations in our Action Research Methodology. 

\subsection{Insights Learned from Community}

By analyzing the feedback from 400 participants of our satisfaction survey, we have convened some engaging and helpful insights. Those who were highly satisfied with our screening events, when asked to identify the critical reasons for satisfaction, they mostly mentioned the comfort of location, involvement, dissemination, and connection. Since the events were arranged in different community locations and during the community events, it was effortless for them to have the long due eye-screening. The Spanish-speaking population mentioned that they felt very involved and connected as they could better communicate with the staffs due to speaking in their native language. The data analysis presented in the evaluation section also concludes that, the ability to communicate in the native language had a very strong impact on the Spanish-speaking population, rather than the rest English-speaking population. 

Moreover, after analyzing the demographic overview of our participant database, we have found that, although our project focused on Latino/Hispanic communities by partnering with Latino Community Centers, it also attracted other communities (28.32\% of our participants belong to African American population). Moreover, when categorized based on insurance, around 40\% of our participants had no health insurance. These demographic concludes about the contribution of our project on diminishing the vision health disparities and about the impact on the communities by contributing to developing better eye-health knowledge. 

\subsection{Future Work}

Our future work includes the integration of a novel Artificial Intelligence (AI) algorithm in our framework to enable automated detection of DR from retinal images \cite{gulshan2016development} as an introductory part of grading. It is expected to help eye-specialist in validating the results. This feature addition would reduce eye-specialists' work burden, which would help them to focus more on the patients who are diagnosed with DR. However, we would conduct an acceptance and feasibility study of this integration before implementing, and we would prioritize the interpretability of the AI algorithm to maintain grading transparency and quality.  
\section{Conclusion}

Proposition of multi-component community focused mHealth tool, and specifically its application in providing eye-health care to the diabetic population has yet not been thoroughly explored in the research community, to our best knowledge. Our approach moves a step forward in that direction, and have shown potential to aid specific communities in getting affordable, accessible, and comfortable health-care support. Our study in this paper provides a more in-depth insight into the problem domain and its related challenges. With the help of community partners, our approach is found to be efficient and covered a large number of population in minimal time. We discussed design considerations to build similar tools and solutions, and through satisfaction survey, we explored insights from the focus community both quantitatively and qualitatively. We expect that our analysis and findings will encourage more researchers to consider these barriers and approaches to remove, both in the field of digital health and human-computer interaction researchers.  
\section{Acknowledgments}
  \label{sec:Acknowledgments}
 
 This work is partially supported by a CTSI grant and funding from the Healthier Wisconsin Partnership Program.

\bibliographystyle{elsarticle-num}
\bibliography{reference}





\end{document}